\documentclass[conference,a4paper]{IEEEtran}

\usepackage{amssymb}
\usepackage{mathrsfs}
\usepackage{graphicx}
\usepackage{color}
\usepackage{url}

\begin{document}

\sloppy

\newtheorem{theorem}{Theorem}
\newtheorem{corollary}{Corollary}
\newtheorem{lemma}{Lemma}
\newtheorem{definition}{Definition}
\newtheorem{claim}{Claim}
\newtheorem{remark}{Remark}
\newtheorem{conjecture}{Conjecture}

\title{Opportunistic Interference Management for Multicarrier systems}

\author{
  \IEEEauthorblockN{Shaunak Mishra}
  \IEEEauthorblockA{UCLA\\
    Email: shaunakmishra@ucla.edu} 
  \and
  \IEEEauthorblockN{I-Hsiang Wang}
  \IEEEauthorblockA{EPFL\\
    Email: i-hsiang.wang@epfl.ch}
  \and
  \IEEEauthorblockN{Suhas Diggavi}
  \IEEEauthorblockA{UCLA\\
    Email: suhasdiggavi@ucla.edu}
}

\maketitle

\begin{abstract}
We study opportunistic interference management when there is bursty interference in parallel $2$-user linear deterministic interference channels. A degraded message set communication problem is formulated to exploit the burstiness of interference in $M$ subcarriers allocated to each user. 
We focus on symmetric rate requirements based on the number of interfered subcarriers rather than the exact set of interfered subcarriers. Inner bounds are obtained using erasure coding, signal-scale alignment and Han-Kobayashi coding strategy. Tight outer bounds for a variety of regimes are obtained using
the El Gamal-Costa injective interference channel 
bounds and a sliding window subset entropy inequality \cite{Tie_liu}. The result demonstrates an application of techniques from multilevel diversity coding to interference channels. We also conjecture outer bounds indicating the sub-optimality of erasure coding across subcarriers in certain regimes.
\end{abstract}
\section{Introduction}
In multicarrier systems like OFDM, subcarriers allocated to a user may face interference due to a variety of reasons. These include the activity of other users and allocation decisions of neighbouring base stations in a cellular network. Predicting the presence or absence of interference in a particular subcarrier may not be feasible at a transmitter in such uncoordinated networks. Nevertheless, it is practical to assume that a subcarrier allocated to a user does not face interference in every channel instantiation. Thus, there is a scope for harnessing such \textit{bursty} interference in multicarrier systems and exploring the possibility of opportunistic rate increments.

\par
\begin{figure}[!ht]
\begin{center}
\includegraphics[scale=0.2]{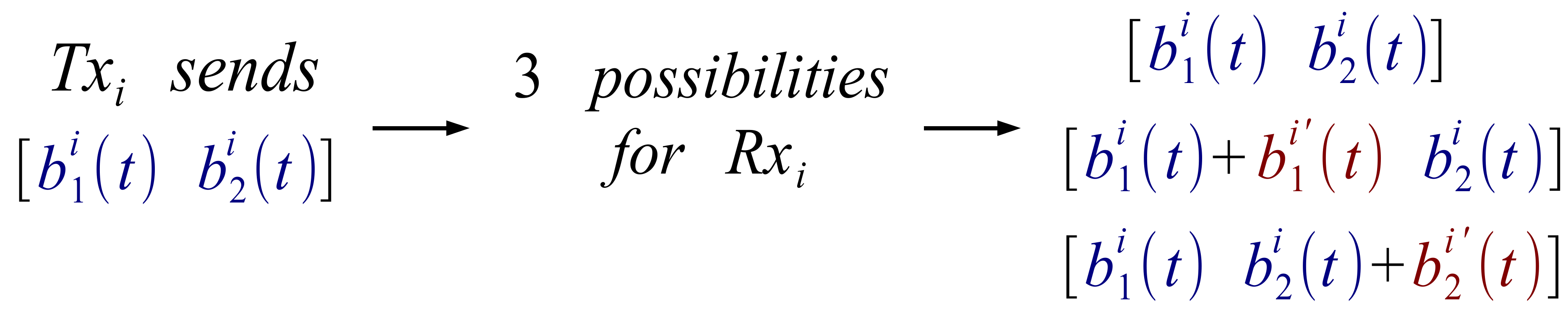}
\caption{Channel realizations for $Rx_i$ in the toy example. The ``$+$'' operator denotes modulo $2$ addition and indicates the presence of interference. As shown above, interference is not present in all channel realizations for $Rx_i$ (hence bursty); but whenever it is present, it is limited to just $1$ out of the $2$ transmitted bits.}
\label{fig:toy_example_wireless}
\end{center}
\end{figure}

 The following toy example, based on parallel linear deterministic channels, captures the intuition behind our problem formulation. Consider $2$ transmitters ($Tx_1$ and $Tx_2$) and $2$ receivers ($Rx_1$ and $Rx_2$). For $i \in \{1,2\}$, $Tx_i$ has messages for $Rx_i$ and at discrete time index $t \in \{1,2,\ldots,N\}$, $Tx_i$ can transmit $2$ bits $[b^i_1(t) \; b^i_2(t)]$. The $2$ bits correspond to $2$ subcarriers (parallel channels) allocated to each transmitter-receiver pair. Depending on the interference channel realization (stays constant for $t \in \{1,2,\ldots,N\}$), $Rx_i$ receives one of the three possibilities: $[ b^i_1(t)\;b^i_2(t)]$, $[b^i_1(t) + b^{i'}_1(t)\;b^i_2(t)]$ and $[ b^i_1(t)\;b^i_2(t) + b^{i'}_2(t)]$ (shown in Figure~\ref{fig:toy_example_wireless}), where $i,i'\in \{1,2\}$ and $i' \neq i$. The first possibility corresponds to the interference free case (for $Rx_i$) and the remaining two possibilities correspond to interference from $Tx_{i'}$ (only one of the subcarriers of $Rx_i$ gets interfered). Hence, there are $3 \times 3 = 9 $ distinct possibilities for the pair of received values at $Rx_1$ and $Rx_2$ over time duration $N$. 
The crucial constraint in this setup is that the transmitters do not know \textit{a priori} the interference channel realization. The channel is used $N$ times (time index $t \in \{1,2,\ldots,N\}$) and we have the following (symmetric) rate requirement: ensure base rate $R_1$ at a receiver when \textit{any} one of the subcarriers (of the receiver) gets interfered and ensure rate $R_0 + R_1$ at a receiver when both subcarriers (of the receiver) are interference free (\emph{i.e.,} opportunistically deliver incremental rate $R_0$, in addition to $R_1$, whenever a receiver is interference free). In this setup, we are interested in characterizing the rate region $(R_1, R_0)$ as the performance metric. Clearly, $R_0 \leq 2$ (a maximum of $2$ bits per time index can be sent by a transmitter) and corner point $(R_1, R_0) = (0,2)$ is easily achievable. Also, the corner point $(R_1, R_0) = (1,0)$ can be easily achieved by using a repetition code across the $2$ subcarriers (\emph{i.e.,} $b^1_1(t) = b^1_2(t)$ and $b^2_1(t) = b^2_2(t)$). The repetition code ensures decodability of the message (of rate $R_1$) irrespective of which subcarrier gets interfered. Using time sharing between corner points $(0,2)$ and $(1,0)$, we can achieve $2R_1 + R_0 \leq 2$. Intuitively this looks like the best we can do, and indeed it can be shown to be tight using entropy inequalities. The problem pursued in this paper is a generalization of this example through parallel linear deterministic interference channels (leading to a rate region with more than two non-trivial corner points in most cases).

\par In \cite{KPV_ISIT09} and \cite{KPV_ITW09}, the problem of harnessing bursty interference was studied for a single carrier scenario using a degraded message set approach. This approach guarantees a base rate when the carrier faces interference. In addition to the base rate, an incremental rate is provided whenever the carrier is interference free. In the multicarrier version considered in this paper, every user (receiver) is allocated $M$ subcarriers (parallel channels) and we extend the degraded message set approach for a rate tuple $(R_0,R_L, R_M)$ as follows: (a) when all $M$ subcarriers of a user get interfered, the user achieves rate $R_M$ (b) when any $L$ out of $M$ subcarriers get interfered, the user achieves rate $R_M + R_L$ and (c) when all $M$ subcarriers are interference free, the user achieves rate $R_M+R_L+R_0$. Thus, the user experiences opportunistic rate increments as the number of interfered subcarriers decreases. Maintaining low message complexity is the practical idea behind considering the number of interfered subcarriers rather than the specific set of subcarriers interfered. The problem formulation has some similarity with symmetric multilevel diversity coding \cite{Roche} and our results demonstrate that similar tools (subset entropy inequalities) as in \cite{Tie_liu} can be used in this context.
\par Our main contributions in this paper are:
\begin{itemize}
\item Inner bounds for $(R_0, R_L, R_M=0)$ and $(R_0=0, R_L, R_M)$ setups using erasure coding across subcarriers (employed for specific interfered \textit{levels} in a subcarrier), signal-scale alignment \cite{KPV_ISIT09,bresler} and Han-Kobayashi scheme.
\item Develop outer bounds using techniques inspired by multilevel diversity codes.
\item The inner and outer bounds coincide for several regimes.
\end{itemize}
The remainder of this paper is organized as follows. Section~\ref{sec:model} formalizes the setup and rate requirements. Section \ref{sec:main_results} states the main results. Inner bounds and outer bounds are discussed in Sections~\ref{sec:inner_bounds} and \ref{sec:outer_bounds} respectively. We conclude the paper with a short discussion in Section~\ref{sec:discussion}.
\section{Notation and setup}\label{sec:model}
We consider a system with two base stations (transmitters) $Tx_1$ and $Tx_2$ and two users (receivers) $Rx_1$ and $Rx_2$. For $i\in \{1,2\}$, user $Rx_i$ is allocated $M$ subcarriers $s^i_1, s^i_2,\ldots s^i_M$ by the base station $Tx_i$. The transmit signals of base stations $Tx_1$ and $Tx_2$ are assumed to be independent. 
\subsection{Channel Model}
The channel is modeled by a $2$-user multicarrier (parallel) linear deterministic interference channel \cite{ADT_det_channel} where, similar to \cite{KPV_ISIT09}, interfering links in each subcarrier may or may not be active (unknown to the transmitters). At discrete time index $t \in \{1,2,\ldots N\}$, the transmit signal on subcarrier $s^i_j$ is $\mathbf{x}^i_j(t) \in \mathbb{F}^q$ where $\mathbb{F}$ is a finite field. The received signals on subcarrier $s^i_j$ of $Rx_i$ when $s^i_j$ faces interference from $s^{i'}_j$ (corresponding to user $i'\neq i $) and when it is interference free are described below as (\ref{eq:interfered}) and (\ref{eq:interference_free}) respectively,
\begin{eqnarray} 
\mathbf{y}^i_j(t)& =& \mathbf{G}^{q-n}\mathbf{x}^i_j(t) + \mathbf{G}^{q-k}\mathbf{x}^{i'}_j(t)  \label{eq:interfered} \\
\mathbf{y}^{i}_j(t) &=& \mathbf{G}^{q-n}\mathbf{x}^i_j(t)   \label{eq:interference_free}
\end{eqnarray}
where $\mathbf{G}$ is a $q \times q$ shift matrix in the terminology of deterministic channel models \cite{ADT_det_channel} and $\mathbf{x}^{i'}_j(t)$ denotes the transmit signal on subcarrier $s^{i'}_j$ for user $i'$. All operations above are in $\mathbb{F}^q$. Similar to \cite{KPV_ISIT09}, the transmitters are assumed to have prior knowledge of parameters $n$ and $k$ (direct and interfering channel strengths), and the presence (or absence) of interference in a subcarrier is assumed to be constant throughout the channel usage duration. Without loss of generality, we assume $q=\max(n,k)$. Let $\alpha = \frac{k}{n}$ denote the normalized strength of the interfering signal. Since interference free capacity for a single carrier can be achieved when $\alpha \geq 2$ \cite{Carleial}, we focus on $0 \leq \alpha \leq 2$. For every time instant, it is convenient to consider a subcarrier as indexed levels of bit pipes. Each bit pipe can carry a symbol from $\mathbb{F}$.
\\Let $ \mathbf{v}^i_j(t)=\mathbf{G}^{q-k}\mathbf{x}^{i'}_j(t) $ denote the interfering signal for $Rx_i$ on subcarrier $s^i_j$. We use $\mathbf{X}^i_j = [\mathbf{x}^i_j(1) \; \mathbf{x}^i_j(2)...\mathbf{x}^i_j(N)]$ to denote the transmit signals sent during $N$ time slots on $s^i_j$ and $\mathbf{V}^i_j$ is defined similarly from $\mathbf{v}^i_j(t)$. Also, we define $\mathbf{X}^i_{j_1:j_2}=[\mathbf{X}^i_{j_1}  \mathbf{X}^i_{j_1 + 1}, \ldots \mathbf{X}^i_{j_2}]$.

\subsection{Rate Requirements}\label{sec:rate_requirements} 
The rate requirements for both the users are constrained to be symmetric.
For $Rx_i$, messages $( W^i_0, W^i_L, W^i_M)$ corresponding to rate tuple $(R^i_0,R^i_L,R^i_M)=(R_0,R_L,R_M)$ are encoded in $\mathbf{X}^i_{1:M}$. Based on the number of interfered subcarriers for $Rx_i$, we have the following 
requirements for the desired messages:
\begin{enumerate}
\item $Rx_i$ decodes $W^i_M$ when all $M$ subcarriers of $Rx_i$ get interfered.
\item $Rx_i$ decodes $(W^i_L,W^i_M)$ when any $L$ out of $M$ subcarriers of $Rx_i$ get interfered.
\item $Rx_i$ decodes $(W^i_0,W^i_L,W^i_M)$ when all $M$ subcarriers of $Rx_i$ are interference free.
\end{enumerate}
A rate tuple is considered achievable if the probability of decoding error is vanishingly small as $N\rightarrow \infty$. To simplify our analysis, we consider two setups: $(R_0,R_L,0)$-setup and $(0,R_L,R_M)$-setup. In the $(R_0,R_L,0)$-setup, $R_M$ is assumed to be zero and in the $(0,R_L,R_M)$-setup $R_0$ is assumed to be zero. The rate regions for these two setups are analyzed separately in this paper.
\section{Main results}  \label{sec:main_results}
Depending on whether $L\leq \frac{M}{2}$ or $L\geq \frac{M}{2}$, we have different results for $(R_0,R_L,0)$-setup and $(0,R_L,R_M)$-setup.
\subsection{Results for $(R_0,R_L,0)$-setup}
\subsubsection{ $L\leq \frac{M}{2}$} We have a tight characterization of capacity in this case.
\begin{theorem} \label{thm:capacity_L_0_less_L} 
For $L\leq \frac{M}{2}$, the capacity region for $(R_0,R_L,0)$-setup  is as follows.
\begin{eqnarray}
M R_L + (M-L)R_0 &\leq& M((M-2L) + L( \max(1,\alpha) \nonumber \\ &&+ \max(1-\alpha, 0)  )  )n \label{eq:OB_1}\\
R_L + R_0 &\leq& Mn \label{eq:OB_2}
\end{eqnarray}
\end{theorem}
\subsubsection{$L\geq \frac{M}{2}$}In this case, we have a tight characterization in certain regimes.
\begin{theorem} \label{thm:inner_L_0_more_L}For $L\geq \frac{M}{2}$, consider the following rate inequalities:
\begin{eqnarray}
M R_L &+& (M-L)R_0  \nonumber \\ &\leq & M( (M-L)(\max(1-\alpha,0) + \max(1,\alpha) )\nonumber \\ &&\quad + (2L-M)\max(\alpha,1-\alpha) )n \label{eq:OB_3} \\
R_L + R_0 &\leq &Mn \label{eq:OB_4}\\
2 R_L  + R_0 &\leq& M( \max(1,\alpha)+ \max(1-\alpha,0) )n \label{eq:OB_5}
\end{eqnarray}
Inequalities (\ref{eq:OB_3}), (\ref{eq:OB_4}) and (\ref{eq:OB_5}) are inner bounds; (\ref{eq:OB_3}) and (\ref{eq:OB_4}) are outer bounds.
\end{theorem}
\begin{corollary} \label{cor:1}
We have a tight characterization for the regime $\{L\geq \frac{M}{2},\; 0\leq \alpha \leq \frac{1}{2}\}$ in the $(R_0,R_L, 0)$-setup. This follows from the observation that (\ref{eq:OB_5}) is not active in presence of (\ref{eq:OB_3}) and (\ref{eq:OB_4}) for $\{L\geq \frac{M}{2},\;0\leq \alpha \leq \frac{1}{2}\} $ (see Appendix for detailed proof).
\end{corollary}
\begin{conjecture} \label{conj:OB_5}
For the $(R_0,R_L,0)$-setup with $L\geq \frac{M}{2}$, (\ref{eq:OB_5}) is an outer bound.
\end{conjecture}
If Conjecture~\ref{conj:OB_5} holds, we have a tight characterization for $(R_0,R_L,0)$-setup when $L\geq \frac{M}{2}$.
\subsection{Results for $(0,R_L,R_M)$-setup}
\subsubsection{ $L\leq \frac{M}{2}$}In this case, we have a tight characterization in certain regimes.
\begin{theorem} \label{thm:inner_M_L_less_L} For $L\leq \frac{M}{2}$, consider the following rate inequalities:
\begin{eqnarray}
R_L + R_M &\leq& ( (M-2L) + L(\max(1,\alpha) \nonumber \\ && + \; \max(1-\alpha,0) ) )n \label{eq:OB_6} \\ 
R_M & \leq & M\max(1-\alpha,\alpha)n  \label{eq:OB_7}\\
M R_L  + 2(M-L) R_M &\leq& M (M-L)( \max(1,\alpha) \nonumber\\ &&+\; \max(1-\alpha,0) )n \label{eq:OB_8}
\end{eqnarray}
Inequalities (\ref{eq:OB_6}), (\ref{eq:OB_7}) and (\ref{eq:OB_8}) are inner bounds; (\ref{eq:OB_6}) and (\ref{eq:OB_7}) are outer bounds.
\end{theorem}
\begin{corollary} \label{cor:2}
We have a tight characterization for the regime $\{ L\leq \frac{M}{2},\;0\leq \alpha \leq \frac{1}{2}\} $ in the $(0,R_L, R_M)$-setup. This follows from the observation that (\ref{eq:OB_8}) is not active in presence of (\ref{eq:OB_6}) and (\ref{eq:OB_7}) for $\{L\leq \frac{M}{2},\;0\leq \alpha \leq \frac{1}{2}\} $ (see Appendix for detailed proof). 
\end{corollary}
\begin{conjecture} \label{conj:OB_8}
For the $(0,R_L,R_M)$-setup with $L\leq \frac{M}{2}$, (\ref{eq:OB_8}) is an outer bound.
\end{conjecture}
If Conjecture~\ref{conj:OB_8} holds, we have a tight characterization for $(0,R_L,R_M)$-setup when $L\leq \frac{M}{2}$.
\subsubsection{ $L\geq \frac{M}{2}$} In this case, we have a tight characterization in certain regimes.
\begin{theorem} \label{thm:inner_M_L_more_L}For $L\geq \frac{M}{2}$, consider the following rate inequalities:
\begin{eqnarray}
R_L + R_M &\leq&  (  (M-L) (  \max(1,\alpha)  +  \max(1-\alpha, 0) ) \nonumber\\ && + (2L-M) \max ( 1-\alpha,\alpha) ) n \label{eq:OB_9}  \\
R_M & \leq & M\max(1-\alpha,\alpha)n  \label{eq:OB_10} \\
 R_L  +  R_M &\leq& \frac{M}{2}  ( \max(1,\alpha) \nonumber\\ &&+ \; \max(1-\alpha,0) )n \label{eq:OB_11}
\end{eqnarray}
Inequalities (\ref{eq:OB_9}), (\ref{eq:OB_10}) and (\ref{eq:OB_11}) are inner bounds; (\ref{eq:OB_9}) and (\ref{eq:OB_10}) are outer bounds.
\end{theorem}
\begin{corollary} \label{cor:3}
We have a tight characterization for the regime $\{L\geq \frac{M}{2},\;0\leq \alpha \leq \frac{2}{3}\} $ in the $(0,R_L, R_M)$-setup. This follows from the observation that (\ref{eq:OB_11}) is not active in presence of (\ref{eq:OB_9}) and (\ref{eq:OB_10}) for $\{L\geq \frac{M}{2},\;0\leq \alpha \leq \frac{2}{3}\} $ (see Appendix for detailed proof).
\end{corollary}
\begin{conjecture} \label{conj:OB_11}
 We conjecture  that (\ref{eq:OB_11}) is an outer bound for $(0,R_L,R_M)$-setup when $L\geq \frac{M}{2}$.
\end{conjecture}
If Conjecture~\ref{conj:OB_11} holds, we have a tight characterization for $(0,R_L,R_M)$-setup when $L\geq \frac{M}{2}$.
\section{Inner bounds} \label{sec:inner_bounds}
Figure~\ref{fig:inner_bounds} summarizes the inner bounds for different regimes depending on values of $\alpha$, $M$ and $L$. The inner bound rate region is obtained from achievable corner points (shown in Figure~\ref{fig:inner_bounds}) using time-sharing. Achievability schemes for corner points shown in Figure~\ref{fig:inner_bounds} can be described as follows.
\subsection{Achievable corner points $(R_L,R_0)$ in $(R_0,R_L,0)$-setup}
\begin{figure*}[!ht]
\begin{center}
\includegraphics[width=7.1in,height=5.45in]{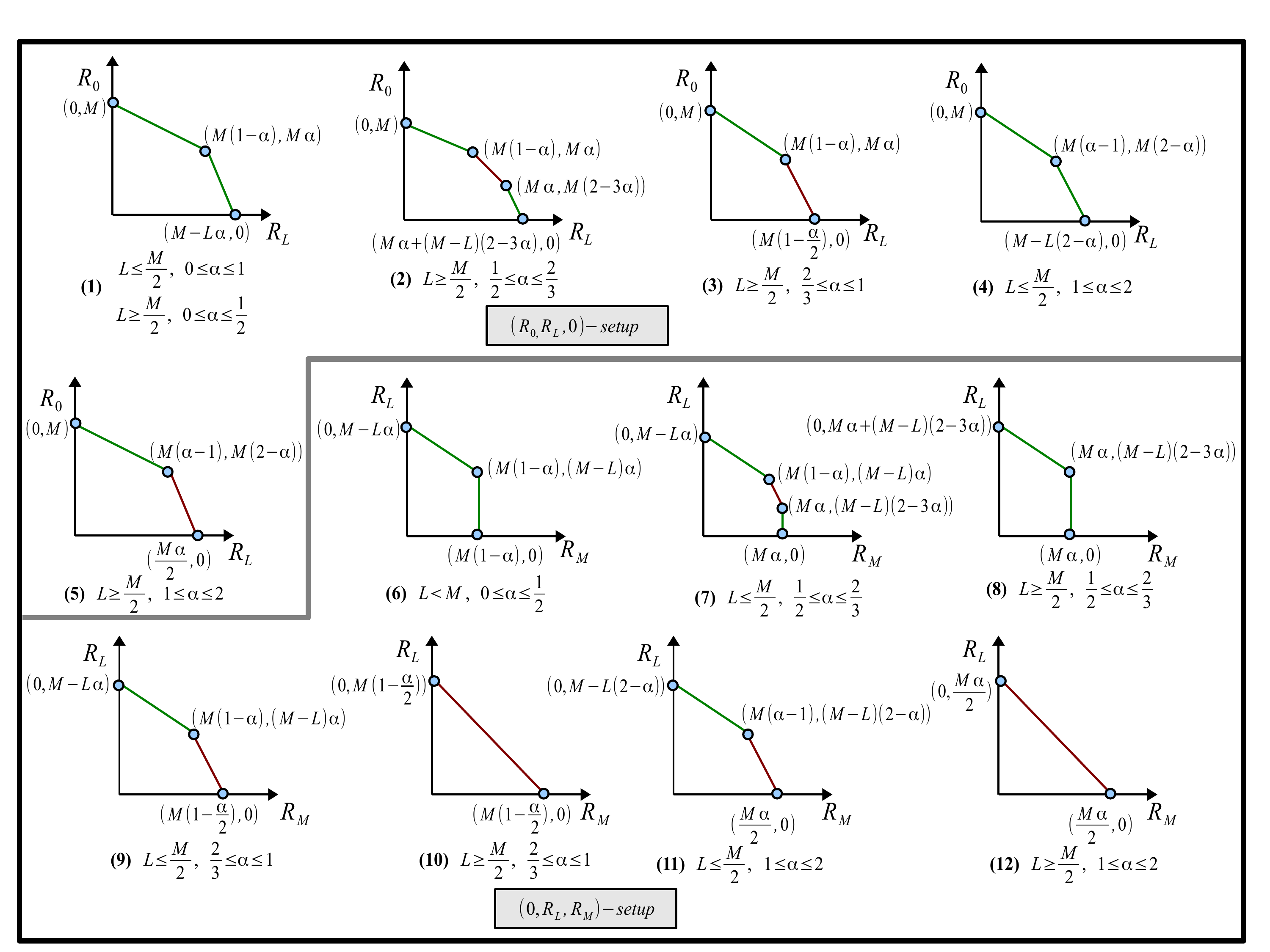}
\caption{Inner bound rate regions for $(0,R_L,R_M)$-setup and $(R_0,R_L,0)$-setup in different regimes. The achievable corner points have been normalized with respect to $n$ and are indicated by blue dots. Lines corresponding to tight outer bounds are colored green and the conjectured outer bounds are colored red.}
\label{fig:inner_bounds}
\end{center}
\end{figure*}

\begin{itemize}
\item \textit{$(0,Mn)$}: This appears in cases (1)-(5) in Figure~\ref{fig:inner_bounds}. It can be achieved by using the top $n$ levels in all the $M$ subcarriers for message $ W^i_0$.
\item \textit{$(M(1-\alpha)n, M\alpha n)$}: This corner point is achievable for $\alpha\leq 1$ and appears in cases (1)-(3) in Figure~\ref{fig:inner_bounds}. To achieve this, the top $(1-\alpha)n$ levels of each subcarrier are used for $W^i_L$ and the bottom $\alpha n$ levels are used for $W^i_0$. Since the top $(1-\alpha) n$ levels of a subcarrier are always interference free, using $M$ subcarriers we achieve $(M(1-\alpha)n, M\alpha n)$.
\item \textit{$((M-L\alpha) n,0)$}: This corner point is achievable for $\alpha\leq 1$ and appears in case (1) in Figure~\ref{fig:inner_bounds}. Since any $L$ out of $M$ subcarriers get interfered, an erasure code\footnote{Interfered levels in the interfered subcarriers are treated as erasures.} (across $M$ subcarriers) can recover symbols at rate $(M-L)\alpha n$ from the bottom $\alpha n $ levels of $M$ subcarriers. Also, an additive rate of $M(1-\alpha)n$ can be obtained by using the top $(1-\alpha)n$ levels of $M$ subcarriers. Adding the contributions from the bottom $\alpha n $ levels  and top $(1-\alpha)n$ levels of all $M$ subcarriers, we achieve $ R_L= (M-L)\alpha n + M (1-\alpha)n  =  (M- L\alpha)n$.
\item \textit{$(M\alpha n , M(2-3\alpha) n)$ and $((M\alpha + (M-L)(2-3\alpha))n, 0)$}: These appear in case (2) in Figure~\ref{fig:inner_bounds} and are achievable for $\frac{1}{2} \leq \alpha\leq \frac{2}{3}$ using the following signal-scale alignment technique \cite{bresler,KPV_ISIT09}. The $n$ levels in a subcarrier $s^i_j$ are divided into $4$ bands $L_1$,$L_2$,$L_3$ and $L_4$ as shown in Figure~\ref{fig:alignment}. For $i\neq i'$, when subcarrier $s^i_j$ faces interference,  only $L_1$ of $s^{i'}_j$ interferes with $L_2$ and $L_3$ of $s^i_j$. Also, only $L_2$ of $s^{i'}_j$ interferes with $L_4$ of $s^{i}_j$. Given this structure, the trick will be to not transmit any information in band $L_2$. This keeps $L_4$ interference free as shown in Figure~\ref{fig:alignment}. Using $L_1$ and $L_4$ of $M$ subcarriers for $ W^i_L$, we achieve $ R_L = M\alpha n$. Using only $L_3$ of $M$ subcarriers for $W^i_0$ we achieve $ R_0 = M(2-3\alpha) n$. Hence $(M\alpha n ,M(2-3\alpha) n)$ is achievable. For $((M\alpha + (M-L)(2-3\alpha))n, 0)$, the same signal-scale alignment trick is used in addition to a rate $\frac{M-L}{M}$ erasure code across $M$ subcarriers for $L_3$. 
\begin{figure}[!ht]
\begin{center}
\includegraphics[width=3.6in,height=1.55in]{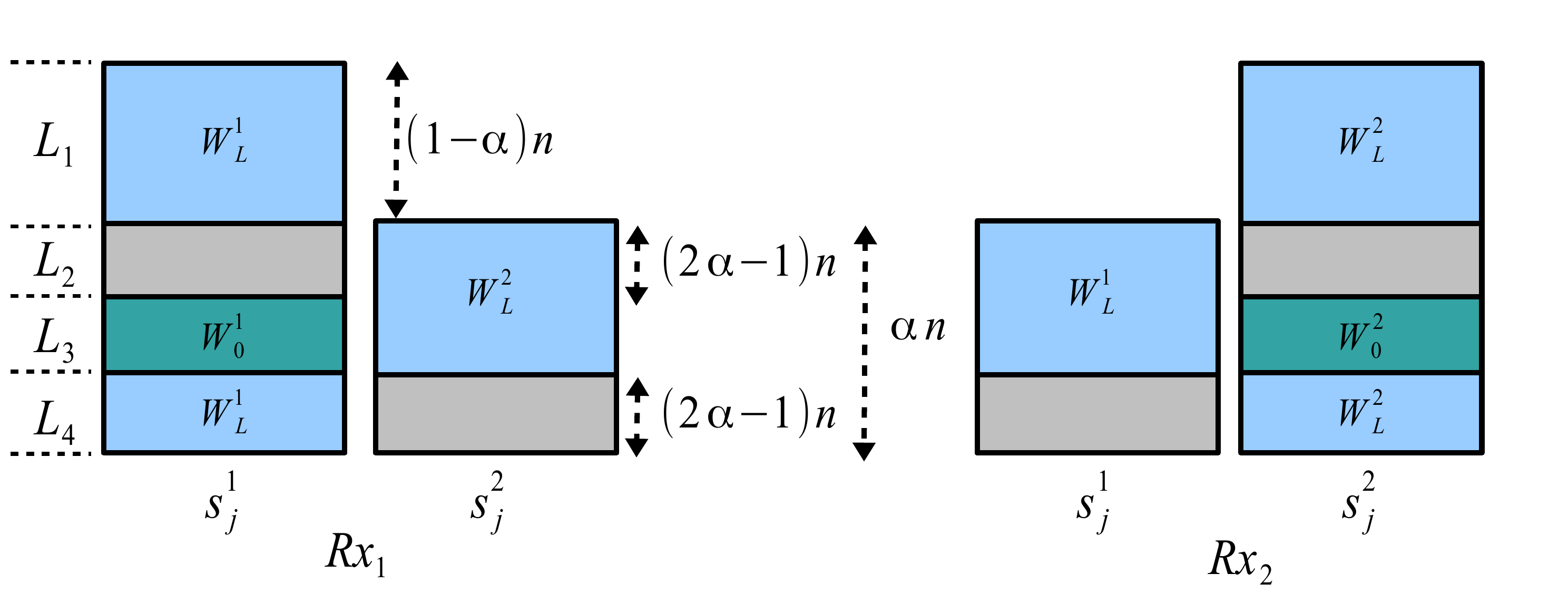}
\caption{Signal-scale alignment technique to achieve $(M\alpha n, M(2-3\alpha)n)$}
\label{fig:alignment}
\end{center}
\end{figure}
\item \textit{$(M(1-\frac{\alpha}{2})n,0)$}: This appears in case (3) in Figure~\ref{fig:inner_bounds} and is achievable for $\frac{2}{3} \leq \alpha \leq 1$. Han-Kobayashi scheme \cite{GK_book} can achieve rate $(1-\frac{\alpha}{2})n$ for a single interfered subcarrier when $\frac{2}{3} \leq \alpha \leq 1$. This scheme is used for each of the $M$ subcarriers to achieve this corner point.
\item \textit{$(M(\alpha-1)n,M(2-\alpha)n)$ and $((M-L(2-\alpha))n,0)$}: These are achievable for $1\leq \alpha \leq 2$. The corner point $(M(\alpha-1)n,M(2-\alpha)n)$ appears in cases (4) and (5) in Figure~\ref{fig:inner_bounds} and is achievable using the following signal-scale alignment strategy. The top $(2-\alpha)n$ levels of a subcarrier are used for $W^i_0$. The next $(\alpha-1)n$ levels are used for $W^i_L$. This ensures that the levels used for $W^i_L$ are always interference free. Using $M$ subcarriers we achieve, $(M(\alpha-1)n,M(2-\alpha)n)$. To achieve $((M-L(2-\alpha))n,0)$ (which appears in case (4) in Figure~\ref{fig:inner_bounds}), a similar scheme is used with a rate $\frac{M-L}{M}$ erasure code (across $M$ subcarriers) for the top $(2-\alpha)n$ levels of a subcarrier.
\item \textit{$(\frac{M \alpha}{2}n,0)$}: This appears in case (5) in Figure~\ref{fig:inner_bounds} and is achievable for $1\leq \alpha \leq 2$. For the classical two user interference channel (single carrier) with $1\leq \alpha \leq 2$, rate $\frac{\alpha}{2}n$ is easily achievable. Using this single carrier scheme for $M$ subcarriers, we achieve $(\frac{M \alpha}{2}n,0)$.
\end{itemize}

\subsection{Achievable corner points $(R_M, R_L)$ in $(0,R_L,R_M)$-setup}
\begin{itemize}
\item\textit{$(M(1-\alpha)n, (M-L)\alpha n)$}: This appears in cases (6), (7) and (9) in Figure~\ref{fig:inner_bounds} and is achievable for $\alpha \leq 1$. Using the top $(1-\alpha)n$ levels of $M$ subcarriers for $W^i_M$, we achieve $R_M=M(1-\alpha)n$. For $W^i_L$, a rate $\frac{M-L}{M}$ erasure code is used for the bottom $\alpha n $ levels across $M$ subcarriers to obtain $R_L = (M-L)\alpha n$.
\item\textit{$(0, (M-L\alpha) n)$}: This appears in cases (6), (7) and (9) in Figure~\ref{fig:inner_bounds} and is achievable for $0 \leq \alpha \leq 1$. The achievability is same as that of $(R_L,R_0)= ((M-L\alpha) n,0)$ in the $(R_0,R_L,0)$-setup.
\item\textit{$(M\alpha n,(M-L)(2-3\alpha)n)$}: This appears in cases (7)-(8) in Figure~\ref{fig:inner_bounds} and is achievable for $\frac{1}{2} \leq \alpha \leq \frac{2}{3}$. A signal-scale alignment technique similar to the one in Figure~\ref{fig:alignment} is used to achieve $R_M=M\alpha n$. Additionally, a rate $\frac{M-L}{M}$ erasure code across $M$ subcarriers for $L_3$ is used to achieve $R_L =(M-L)(2-3\alpha)n$.
\item\textit{$(0,(M\alpha + (M-L)(2-3\alpha) ) n)$}: This appears in case (8) in Figure~\ref{fig:inner_bounds} and is achievable for $\frac{1}{2} \leq \alpha \leq \frac{2}{3}$. The achievability is same as that of $(R_L,R_0)= ((M\alpha  + (M-L)(2-3\alpha) )n,0)$ in the $(R_0,R_L,0)$-setup.
\item\textit{$(0, M(1-\frac{\alpha}{2})n)$ and $(M(1-\frac{\alpha}{2})n,0)$}: The corner point $(M(1-\frac{\alpha}{2})n,0)$ appears in cases (9) and (10) while $(0, M(1-\frac{\alpha}{2})n)$ appears in case (10) in Figure~\ref{fig:inner_bounds}. Both corner points are achievable for $\frac{2}{3} \leq \alpha \leq 1$. To achieve $(0, M(1-\frac{\alpha}{2})n)$, we use the scheme for achieving $(R_L,R_0) = (M(1-\frac{\alpha}{2})n,0)$ in the $(R_0,R_L,0)$-setup (\emph{i.e.}, Han-Kobayashi scheme is used for all the $M$ subcarriers). Also, by using $W^i_M$ instead of $W^i_L$, the above scheme achieves corner point $(M(1-\frac{\alpha}{2})n,0)$ in the $(0,R_L,R_M)$-setup.
\item \textit{$(M(\alpha-1)n, (M-L)(2-\alpha)n)$ and $(0,(M-L(2-\alpha))n)$}: These are achievable for $1\leq \alpha \leq 2$. The corner point $(M(\alpha-1)n,(M-L)(2-\alpha)n)$ appears in case (11) in Figure~\ref{fig:inner_bounds} and is achievable using the following signal-scale alignment strategy. The top $(2-\alpha)n$ levels of a subcarrier are used for $W^i_L$ with a rate $\frac{M-L}{M}$ erasure code across $M$ subcarriers. The next $(\alpha-1)n$ levels are used for $W^i_M$. This ensures that the levels used for $W^i_M$ are always interference free. Using $M$ subcarriers we achieve, $(M(\alpha-1)n,(M-L)(2-\alpha)n)$. To achieve $(0, M-L(2-\alpha))n,0)$ (which appears in case (11) in Figure~\ref{fig:inner_bounds}), we use the same scheme as that for $(R_L,R_0)= ( (M-L(2-\alpha) ) n, 0 )$ in the $(R_0,R_L,0)$-setup. 
\item\textit{$(0, \frac{M\alpha}{2}n)$ and $(\frac{M\alpha}{2}n,0)$}: The corner point $(\frac{M\alpha}{2}n,0)$ appears in cases (11) and (12) while $(0,\frac{M\alpha}{2}n)$ appears in case (12) in Figure~\ref{fig:inner_bounds}. Both corner points are achievable for $1 \leq \alpha \leq 2$. To achieve $(0, \frac{M\alpha}{2}n)$, we use the scheme for achieving $(R_L,R_0) = (\frac{M \alpha}{2}n,0)$ in the $(R_0,R_L,0)$-setup (case(5) in Figure~\ref{fig:inner_bounds}). Also, by using $W^i_M$ instead of $W^i_L$, the above scheme achieves the corner point $(\frac{M\alpha}{2}n,0)$ in the $(0,R_L,R_M)$-setup.

\end{itemize}
\section{Outer Bounds} \label{sec:outer_bounds}
In this section, we first define additional notation for outer bound proofs. This is followed by outer bound proofs for $(R_0,R_L,0)$-setup (which use techniques \cite{Tie_liu} from multilevel diversity coding) and outer bound proofs for $(0,R_L,R_M)$-setup.
\subsection{Receiver Configurations}
There are $M \choose L$ ways in which any $L$ out of $M$ subcarriers get interfered. Every such choice is a receiver configuration for a user. We use additional notation for a special set of receiver configurations described below. Consider a circulant matrix $\mathbf{C}_{M,L}$ of dimension $M$ with the first row consisting of $M-L$ consecutive ones followed by $L$ zeros. The other rows are cyclic right shifts of the first row. As an example, $\mathbf{C}_{3,1}$ is shown below.
\begin{eqnarray}
\mathbf{C}_{3,1}=\left( \matrix{1&1&0 \cr 0&1&1  \cr 1&0&1}\right) \nonumber
\end{eqnarray} 
\\We use $\mathbf{C}_{M,L}$ to list a specific set of receiver configurations in the following manner. Each row corresponds to a receiver configuration with $M$ subcarriers indexed by the columns. In each row, $1$ denotes an interference free subcarrier and $0$ denotes an interfered subcarrier. Hence, out of $M \choose L$ choices, $\mathbf{C}_{M,L}$ lists only $M$ receiver configurations. For example, the third row in $\mathbf{C}_{3,1}$ shown above indicates a situation for $Rx_i$ where only subcarrier $s^i_2$ gets interfered. The structure of $C_{M,L}$  corresponds to the choice of receiver configurations we use in some of our outer bound proofs. This structure enables the use of sliding window subset inequality \cite{Tie_liu} in such proofs.
\par We now describe additional notation related to receiver configurations of a user. When $Rx_i$ is in receiver configuration indicated by row $j$ of $\mathbf{C}_{M,L}$,  we use $\mathcal{Y}^{i}_{M,L,j}$ to denote the received signal on $M$ subcarriers (over $N$ time slots). In the same spirit, we define $\mathcal{V}^{i}_{M,L,j}$ as the interfering signal over all $M$ subcarriers for $Rx_i$ in this receiver configuration. The received signal in interference free subcarriers in $\mathcal{Y}^{i}_{M,L,j}$ is denoted by $\mathcal{X}^i_{M,L,j}$ and the received signal in interfered subcarriers in $\mathcal{Y}^{i}_{M,L,j}$ is denoted by $\tilde{\mathcal{Y}}^i_{M,L,j}$. When all $M$ subcarriers of $Rx_i$ are interference free, the received signal is denoted by $\mathcal{X}^i_{M,0}=\mathcal{X}^i_{M,0,j}$. \par Now, a direct consequence of the sliding window subset inequality \cite{Tie_liu} in our setting can be stated as follows.
\begin{eqnarray}
\sum_{j=1}^{M} H(\mathcal{X}^i_{M,M-1,j}) &\geq& \frac{1}{2} \sum_{j=1}^{M} H(\mathcal{X}^i_{M,M-2,j}) \ldots \nonumber\\ \ldots &\geq& \frac{1}{M} \sum_{j=1}^{M} H(\mathcal{X}^i_{M,0,j}) \label{eq:sliding_window}
\end{eqnarray}
\subsection{Outer bounds for $(R_0,R_L,0)$-setup}
\subsubsection{Proof of outer bound (\ref{eq:OB_1})}
We prove outer bound (\ref{eq:OB_1}) using a careful choice of receiver configurations represented by rows of $\mathbf{C}_{M,L}$. The high level idea is to divide the received signal into interfered and interference free terms followed by the use of (\ref{eq:sliding_window}) on the interference free terms. The proof can be described as follows.\par Using Fano's inequality for $Rx_i$ $i \in \{1,2\}$, for any $\epsilon > 0$ there exists a large enough $N$ such that,
\begin{eqnarray}
&&N(M  R_L + (M-L) R_0 -(2M-L)\epsilon)\nonumber\\
&\leq &   (M-L) I( W^i_0; \mathcal{X}^i_{M,0} |  W^i_L)\nonumber\\ &&+\quad
\sum_{j=1}^M I( W^i_L; \mathcal{X}^i_{M,L,j} \tilde{\mathcal{Y}}^i_{M,L,j}) \nonumber\\
&=&(M-L)H(\mathcal{X}^i_{M,0} |  W^i_L)  - \sum_{j=1}^M H (\mathcal{X}^i_{M,L,j} | W^i_L)   \nonumber\\ &&+ \sum_{j=1}^M H (\mathcal{X}^i_{M,L,j}) +   \sum_{j=1}^M I( W^i_L;\tilde{\mathcal{Y}}^i_{M,L,j}| \mathcal{X}^i_{M,L,j}) \nonumber\\
&\stackrel{(a)}\leq &\sum_{j=1}^M H (\mathcal{X}^i_{M,L,j})  +   \sum_{j=1}^M I( W^i_L;\tilde{\mathcal{Y}}^i_{M,L,j}| \mathcal{X}^i_{M,L,j}) \label{eq:A_00}\\
&\leq& \sum_{j=1}^M H (\mathcal{X}^i_{M,L,j}) +   \sum_{j=1}^M H(\tilde{\mathcal{Y}}^i_{M,L,j})  \nonumber\\ &&- \sum_{j=1}^M H(\tilde{\mathcal{Y}}^i_{M,L,j}| \mathcal{X}^i_{M,L,j}  W^i_L  W^i_0) \nonumber\\
&= &\sum_{j=1}^M H (\mathcal{X}^i_{M,L,j})  \nonumber\\&&+   \sum_{j=1}^M H(\tilde{\mathcal{Y}}^i_{M,L,j}) - \sum_{j=1}^M H(\mathcal{V}^{i}_{M,L,j} ) \label{eq:A_0}\\
&\stackrel{(b)}\leq &\frac{M-L}{L}\sum_{j=1}^M H (\mathcal{X}^i_{M,M-L,j} )  \nonumber\\ &&+   \sum_{j=1}^M H(\tilde{\mathcal{Y}}^i_{M,L,j}) - \sum_{j=1}^M H(\mathcal{V}^{i}_{M,L,j} )  \label{eq:A_1}
\end{eqnarray}
(a) follows from (\ref{eq:sliding_window}) and (b) follows from $M-L \geq L $ and (\ref{eq:sliding_window}).
Substituting $i=1$ and $i=2$ in (\ref{eq:A_1}), we can obtain two inequalities corresponding to different users. On adding these two inequalities,
\begin{eqnarray}
&&2N(M  R_L + (M-L) R_0 -(2M-L)\epsilon)\nonumber\\
&\leq&\frac{(M-2L)}{L}\sum_{j=1}^M H (\mathcal{X}^1_{M,M-L,j})  +   \sum_{j=1}^M H(\tilde{\mathcal{Y}}^1_{M,L,j}) \nonumber\\
&&+\frac{(M-2L)}{L}\sum_{j=1}^M H (\mathcal{X}^2_{M,M-L,j}) +    \sum_{j=1}^M H(\tilde{\mathcal{Y}}^2_{M,L,j})  \nonumber\\
&&+  ( \sum_{j=1}^M H (\mathcal{X}^1_{M,M-L,j})    - \sum_{j=1}^M H(\mathcal{V}^{2}_{M,L,j}) ) \nonumber\\
&&+   ( \sum_{j=1}^M H (\mathcal{X}^2_{M,M-L,j})  - \sum_{j=1}^M H(\mathcal{V}^{1}_{M,L,j} ) ) \nonumber\\
&\stackrel{(a)}\leq&2N(  M(M-2L)  \nonumber\\ && + ML\max(1,\alpha) + ML\max(1-\alpha ,0) ) n
\end{eqnarray} 
where (a) follows from the structure of $\mathbf{C}_{M,L}$.
\subsubsection{Proof of outer bound (\ref{eq:OB_3})}
\noindent The inequality (\ref{eq:A_00})  in the proof of outer bound (\ref{eq:OB_1}) also holds for $L\geq \frac{M}{2}$. Hence,
\begin{eqnarray}
&&N(M  R_L + (M-L) R_0 -(2M-L)\epsilon)\nonumber\\
&\leq &\sum_{j=1}^M H (\mathcal{X}^i_{M,L,j})  +   \sum_{j=1}^M I( W^i_L;\tilde{\mathcal{Y}}^i_{M,L,j}| \mathcal{X}^i_{M,L,j}) \nonumber \\
&\leq& \sum_{j=1}^M H(\mathcal{X}^i_{M,L,j} \tilde{\mathcal{Y}}^i_{M,L,j}) - \sum_{j=1}^M H(\mathcal{V}^{i}_{M,L,j}) \label{eq:A_10}
\end{eqnarray}
Substituting $i=1$ and $i=2$ in (\ref{eq:A_10}), we can obtain two inequalities corresponding to different users. On adding these two inequalities,
\begin{eqnarray}
&&2N(M R_L + (M-L) R_0 -(2M-L)\epsilon)\nonumber\\
&\leq &\sum_{j=1}^M H(\mathcal{X}^1_{M,L,j} \tilde{\mathcal{Y}}^1_{M,L,j}) - \sum_{j=1}^M H(\mathcal{V}^{2}_{M,L,j})\nonumber\\ && + \sum_{j=1}^M H(\mathcal{X}^2_{M,L,j} \tilde{\mathcal{Y}}^2_{M,L,j}) - \sum_{j=1}^M H(\mathcal{V}^{1}_{M,L,j}) \nonumber\\
&\stackrel{(a)}\leq&2N( M(M-L)(\max(1-\alpha,0) + \max(1,\alpha) )\nonumber \\ &&\quad + M(2L-M)\max(\alpha,1-\alpha) ) n
\end{eqnarray} 
where (a) follows from $L\geq \frac{M}{2}$ and the structure of $\mathbf{C}_{M,L}$.
\subsection{Outer bounds for $(0,R_L,R_M)$-setup}
Outer bounds (\ref{eq:OB_6}), (\ref{eq:OB_7}), (\ref{eq:OB_9}) and (\ref{eq:OB_10}) can be shown by using the El Gamal-Costa injective interference channel bounds \cite{GK_book} as follows. 
\subsubsection{Proof of outer bound (\ref{eq:OB_6})}
For this outer bound proof, we consider two receiver configurations with no interfered subcarriers in common and apply the injective channel bound \cite{GK_book} as shown below.
\par For any $\epsilon > 0$ there exists a large enough $N$ such that,
\begin{eqnarray}
&& N(2 (R_M + R_L)- 2\epsilon) \nonumber\\ 
&\stackrel{(a)}\leq& H(\mathcal{Y}^1_{M,L,1}| \mathcal{V}^2_{M, L ,L+1}) + H(\mathcal{Y}^2_{M,L,L+1}| \mathcal{V}^1_{M,L,1}) \nonumber\\
&\leq& 2N ( L (\max (1, \alpha) \nonumber \\ && +\max(1-\alpha,0) ) + M-2L)n
\end{eqnarray}
where (a) follows from the injective channel bound \cite{GK_book}.
\subsubsection{Proof of outer bounds (\ref{eq:OB_7}) and (\ref{eq:OB_10})}
Outer bounds (\ref{eq:OB_7}) and (\ref{eq:OB_10}) have the same proof. For the proof, we consider the receiver configuration with all $M$ subcarriers interfered and apply the injective channel bound \cite{GK_book} as shown below. 
\par For any $\epsilon > 0$ there exists a large enough $N$ such that,
\begin{eqnarray}
&& N(2 R_M - 2\epsilon) \nonumber\\ 
&\stackrel{(a)}\leq& H(\mathcal{Y}^1_{M,M,j}| \mathcal{V}^2_{M,M,j}) + H(\mathcal{Y}^2_{M,M,j}| \mathcal{V}^1_{M,M,j}) \nonumber\\
&\leq& 2N M \max (1-\alpha, \alpha)n 
\end{eqnarray}
where $\mathcal{Y}^i_{M,M,j}$ corresponds to the receiver configuration with all $M$ subcarriers interfered and (a) follows from the injective channel bound \cite{GK_book}.
\subsubsection{Proof of outer bound (\ref{eq:OB_9})}
For this outer bound proof, we consider two receiver configurations with minimum number of interfered subcarriers (\emph{i.e.,} $2L-M$) in common and apply the injective channel bound \cite{GK_book} as shown below.
\par For any $\epsilon > 0$ there exists a large enough $N$ such that,
\begin{eqnarray}
&& N(2 (R_M + R_L)- 2\epsilon) \nonumber\\ 
&\stackrel{(a)}\leq& H(\mathcal{Y}^1_{M,L,1}| \mathcal{V}^2_{M, L ,L+1}) + H(\mathcal{Y}^2_{M,L,L+1}| \mathcal{V}^1_{M,L,1}) \nonumber\\
&\leq& 2N ( (M- L) (\max (1, \alpha) +\max(1-\alpha,0) )  \nonumber \\ &&+\; (2L-M) \max(1-\alpha, \alpha) ) n
\end{eqnarray}
where (a) follows from the injective channel bound \cite{GK_book}.
\section{Discussion}\label{sec:discussion}
It is optimal to treat interference as noise in the regimes where erasure coding across subcarriers leads to tight inner bounds. However, outer bound conjectures on (\ref{eq:OB_5}) and (\ref{eq:OB_11}) (\emph{i.e.,} Conjectures \ref{conj:OB_5} and \ref{conj:OB_11}) suggest that this may not be the case for all regimes. For $\alpha=1$ (and $L > \frac{M}{2}$), both imply $R_L \leq \frac{M}{2}n$; this can be simply achieved by dividing the $M$ subcarriers between the two users. An erasure coding scheme in this case will lead to $R_L = (M-L)n < \frac{M}{2}n$. Hence, erasure coding across subcarriers may not be optimal in all regimes.

\section*{Acknowledgment}
The work of S. Mishra and S. Diggavi was supported in part by NSF award 1136174 and MURI award AFOSR FA9550-09-064. The work of I.-H. Wang was supported by EU project CONECT FP7-ICT-2009-257616.

\appendix \label{sec:appendix}
\subsection{Proof of Corollary~\ref{cor:1}}
For $\{L\geq \frac{M}{2},\; 0\leq \alpha \leq \frac{1}{2}\}$, inequality (\ref{eq:OB_5}) is not active in presence of inequalities (\ref{eq:OB_3}) and (\ref{eq:OB_4}). This can be proved as follows.

\par
\begin{figure}[!ht]
\begin{center}
\includegraphics[width=3.0in,height=2.55in]{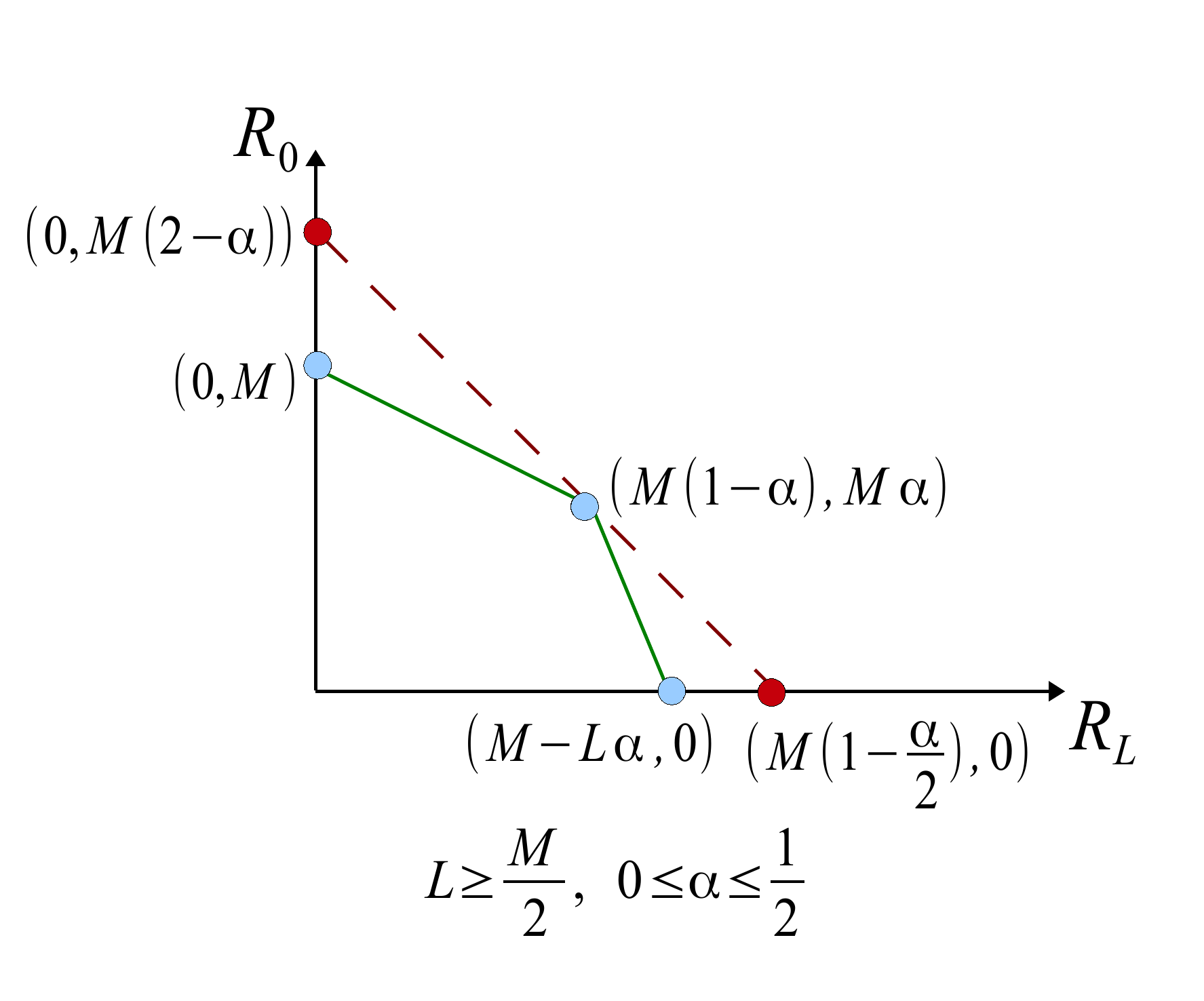}
\caption{Rate inequalities (normalized with respect to $n$) for the regime $\{L\geq \frac{M}{2},\; 0\leq \alpha \leq \frac{1}{2}\}$ in the $(R_0,R_L,0)$-setup. Inequality (\ref{eq:simp_OB_5}) (dashed red line) is not active in presence of (\ref{eq:simp_OB_3}) and (\ref{eq:simp_OB_4}) (solid green lines).}
\label{fig:cor_1}
\end{center}
\end{figure}

 In this regime, inequalities (\ref{eq:OB_3}), (\ref{eq:OB_4}) and (\ref{eq:OB_5}) can be rewritten (shown below) as (\ref{eq:simp_OB_3}), (\ref{eq:simp_OB_4}) and (\ref{eq:simp_OB_5}) respectively.
\begin{eqnarray}
M R_L + (M-L) R_0 &\leq& M(M-L\alpha)n \label{eq:simp_OB_3} \\
R_L + R_0 &\leq &Mn \label{eq:simp_OB_4}\\
2 R_L + R_0 &\leq& M(2-\alpha)n \label{eq:simp_OB_5}
\end{eqnarray}
Figure~\ref{fig:cor_1} shows the situation in this regime\footnote{For $L \geq \frac{M}{2}$, $M-L\alpha = M(1-\frac{\alpha}{2}) + (\frac{M}{2}-L)\alpha \leq  M(1-\frac{\alpha}{2})$ }; it is clear that (\ref{eq:simp_OB_5}) (dashed red line in Figure~\ref{fig:cor_1}) is not active in presence of (\ref{eq:simp_OB_3}) and (\ref{eq:simp_OB_4}) (solid green lines in Figure~\ref{fig:cor_1}). Since inequalities (\ref{eq:simp_OB_3}) and (\ref{eq:simp_OB_4}) are inner bounds as well as outer bounds in this regime, we have a tight characterization.
\subsection{Proof of Corollary~\ref{cor:2}}
For $\{L\leq \frac{M}{2},\; 0\leq \alpha \leq \frac{1}{2}\}$, inequality (\ref{eq:OB_8}) is not active in presence of inequalities (\ref{eq:OB_6}) and (\ref{eq:OB_7}). This can be proved as follows.
\par
\begin{figure}[!ht]
\begin{center}
\includegraphics[width=3.0in,height=2.43in]{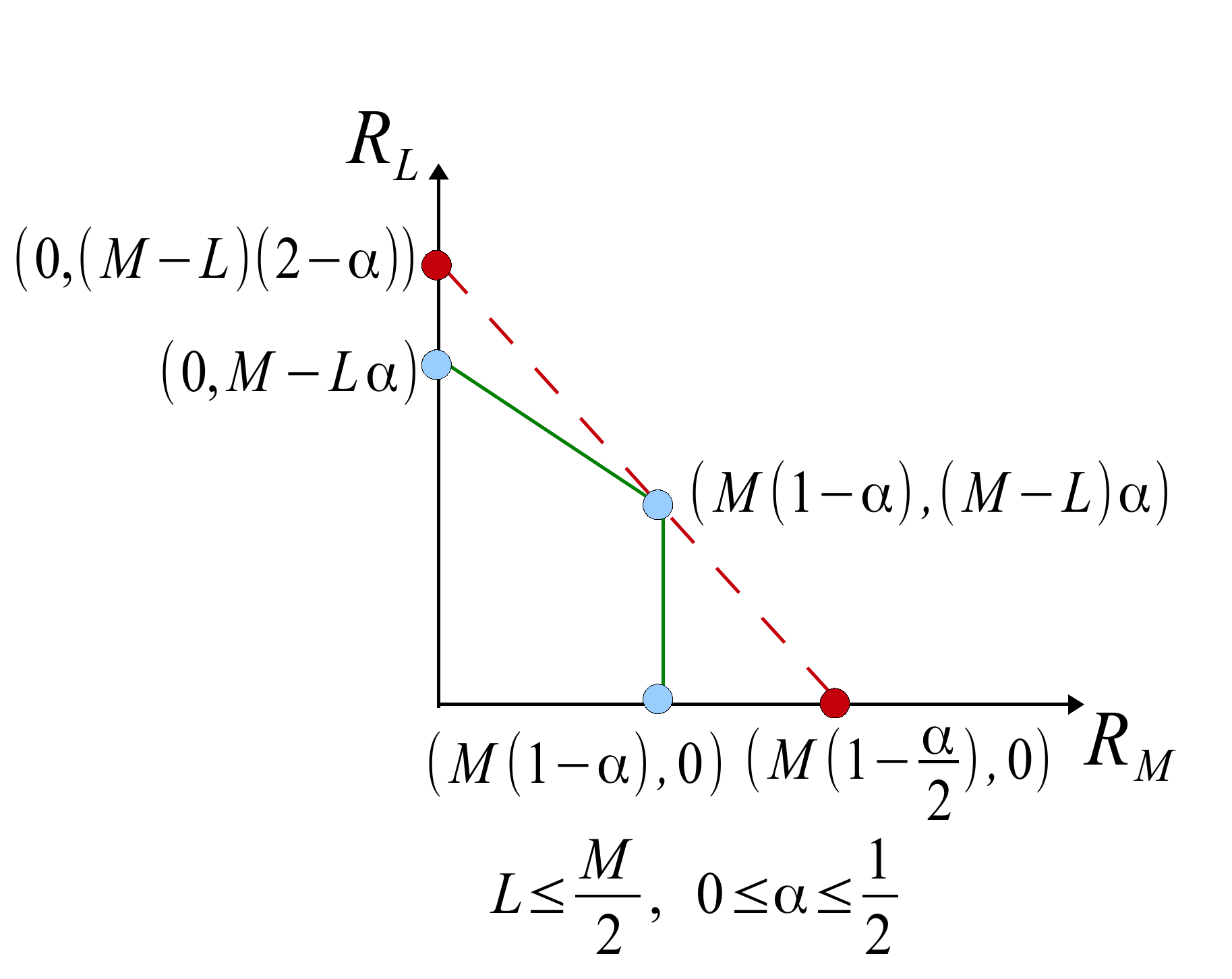}
\caption{Rate inequalities (normalized with respect to $n$) for the regime $\{L\leq \frac{M}{2},\; 0\leq \alpha \leq \frac{1}{2}\}$ in the $(0,R_L,R_M)$-setup. Inequality (\ref{eq:simp_OB_8}) (dashed red line) is not active in presence of (\ref{eq:simp_OB_6}) and (\ref{eq:simp_OB_7}) (solid green lines).}
\label{fig:cor_2}
\end{center}
\end{figure}

 In this regime, inequalities (\ref{eq:OB_6}), (\ref{eq:OB_7}) and (\ref{eq:OB_8}) can be rewritten (shown below) as (\ref{eq:simp_OB_6}), (\ref{eq:simp_OB_7}) and (\ref{eq:simp_OB_8}) respectively.
\begin{eqnarray}
R_L + R_M &\leq& ( M - L \alpha )n \label{eq:simp_OB_6} \\ 
R_M & \leq & M(1-\alpha)n  \label{eq:simp_OB_7}\\
M R_L  + 2(M-L) R_M &\leq& M (M-L)(2-\alpha)n \label{eq:simp_OB_8}
\end{eqnarray}
Figure~\ref{fig:cor_2} shows the situation in this regime; it is clear that (\ref{eq:simp_OB_8}) (dashed red line in Figure~\ref{fig:cor_2}) is not active in presence of (\ref{eq:simp_OB_6}) and (\ref{eq:simp_OB_7}) (solid green lines in Figure~\ref{fig:cor_2}). Since inequalities (\ref{eq:simp_OB_6}) and (\ref{eq:simp_OB_7}) are inner bounds as well as outer bounds in this regime, we have a tight characterization.
\subsection{Proof of Corollary~\ref{cor:3}}
In the regime $\{L\geq \frac{M}{2},\;0\leq \alpha \leq \frac{2}{3}\} $, inequality (\ref{eq:OB_11}) is not active in presence of inequalities (\ref{eq:OB_9}) and (\ref{eq:OB_10}). This can be shown as follows.
\par For this regime, inequalities (\ref{eq:OB_9}), (\ref{eq:OB_10}) and (\ref{eq:OB_11}) can be rewritten (shown below) as (\ref{eq:simp_OB_9}), (\ref{eq:simp_OB_10}) and (\ref{eq:simp_OB_11}) respectively.
\begin{eqnarray}
R_L + R_M &\leq&  (  (M-L) ( 2- \alpha) \nonumber \\ && + (2L-M) \max ( 1-\alpha,\alpha) ) n \label{eq:simp_OB_9}  \\
R_M & \leq & M\max(1-\alpha,\alpha)n  \label{eq:simp_OB_10} \\
R_L  +  R_M &\leq& \frac{M}{2}  ( 2 -\alpha )n \label{eq:simp_OB_11}
\end{eqnarray}
To show (\ref{eq:simp_OB_11}) is not active in presence of (\ref{eq:simp_OB_9}) and (\ref{eq:simp_OB_10}), it is sufficient to prove (\ref{eq:simp_OB_9}) \textit{dominates}\footnote{gives a smaller bound for $R_L + R_M $} (\ref{eq:simp_OB_11}) in this regime. We prove this in two steps as shown below (analysis for $0\leq \alpha \leq \frac{1}{2}$ followed by analysis for $\frac{1}{2}\leq \alpha \leq \frac{2}{3}$).
\par For $0\leq \alpha \leq \frac{1}{2}$, (\ref{eq:simp_OB_9}) can be simplified to
\begin{eqnarray}
R_L + R_M &\leq&  ( M - L \alpha)n \nonumber 
\end{eqnarray}
Since $L\geq \frac{M}{2}$;  $( M - L \alpha) \leq \frac{M}{2}  ( 2 -\alpha )$. Thus, (\ref{eq:simp_OB_9}) dominates (\ref{eq:simp_OB_11}) for $\{L\geq \frac{M}{2},\;0\leq \alpha \leq \frac{1}{2}\}$.

\par For $\frac{1}{2}\leq \alpha \leq \frac{2}{3}$, (\ref{eq:simp_OB_9}) can be simplified to
\begin{eqnarray}
R_L + R_M &\leq&  ( M(2-2\alpha) - L ( 2- 3\alpha) ) n  \nonumber\\
&=& ( \frac{M}{2}(2-\alpha) +  (\frac{M}{2}-L)( 2- 3\alpha))n\nonumber
\end{eqnarray}
Since $ \alpha \leq \frac{2}{3}$ and $\frac{M}{2} \leq L$, $\frac{M}{2}(2-\alpha) +  (\frac{M}{2}-L)( 2- 3\alpha) \leq \frac{M}{2}  ( 2 -\alpha )$. Thus, (\ref{eq:simp_OB_9}) dominates (\ref{eq:simp_OB_11}) for $\{L\geq \frac{M}{2},\;\frac{1}{2}\leq \alpha \leq \frac{2}{3}\}$. 

\par As shown above, (\ref{eq:simp_OB_9}) dominates (\ref{eq:simp_OB_11}) for both $\{L\geq \frac{M}{2},\;0\leq \alpha \leq \frac{1}{2}\}$ and $\{L\geq \frac{M}{2},\;\frac{1}{2}\leq \alpha \leq \frac{2}{3}\}$. Since inequalities (\ref{eq:simp_OB_9}) and (\ref{eq:simp_OB_10}) are inner bounds as well as outer bounds, we have a tight characterization in the regime $\{L\geq \frac{M}{2},\;0\leq \alpha \leq \frac{2}{3}\}$.

\end{document}